\documentclass[runningheads]{llncs}

\usepackage{graphicx}
\usepackage{booktabs}

\usepackage{xcolor}
\usepackage{cite}
\usepackage{subcaption}

\usepackage{amsmath}

\usepackage{amssymb}
\usepackage{multirow}

\usepackage[colorlinks=true, urlcolor=blue, linkcolor=red]{hyperref}

\begin{document}
%
% \title{Cardiovascular Disease Detection from Two Views of Chest X-ray with Mamba}
\title{Cardiovascular Disease Detection from Multi-View Chest X-rays with BI-Mamba}
\titlerunning{CVD Detection from chest X-rays with BI-Mamba}
\author{Zefan Yang\inst{1} \and Jiajin Zhang\inst{1} \and Ge Wang\inst{1} \and \\Mannudeep K. Kalra\inst{2} \and Pingkun Yan\inst{1}}
\authorrunning{Zefan Yang et al.}
\institute{Department of Biomedical Engineering and Center for Biotechnology and Interdisciplinary Studies, Rensselaer Polytechnic Institute, Troy, NY, USA \and Department of Radiology, Massachusetts General Hospital, Harvard Medical School, Boston, MA, USA}
\maketitle              % typeset the header of the contribution
\begin{abstract}
% Cardiovascular diseases (CVDs) are the leading cause of death globally. 
Accurate prediction of Cardiovascular disease (CVD) risk in medical imaging is central to effective patient health management. 
Previous studies have demonstrated that imaging features in computed tomography (CT) can help predict CVD risk. However, CT entails notable radiation exposure, which may result in adverse health effects for patients. In contrast, chest X-ray emits significantly lower levels of radiation, offering a safer option. This rationale motivates our investigation into the feasibility of using chest X-ray for predicting CVD risk.
Convolutional Neural Networks (CNNs) and Transformers are two established network architectures for computer-aided diagnosis. However, they struggle to model very high resolution chest X-ray due to the lack of large context modeling power or quadratic time complexity. Inspired by state space sequence models (SSMs), a new class of network architectures with competitive sequence modeling power as Transfomers and linear time complexity, we propose Bidirectional Image Mamba (BI-Mamba) to complement the unidirectional SSMs with opposite directional information. BI-Mamba utilizes parallel forward and backwark blocks to encode longe-range dependencies of multi-view chest X-rays. We conduct extensive experiments on images from 10,395 subjects in National Lung Screening Trail (NLST). Results show that BI-Mamba outperforms ResNet-50 and ViT-S with comparable parameter size, and saves significant amount of GPU memory during training. Besides, BI-Mamba achieves promising performance compared with previous state of the art in CT, unraveling the potential of chest X-ray for CVD risk prediction.
% Background of CVD -> Disadvantages of LDCT, Advantages of Chest X-ray, Chest X-ray for CVD risk prediction -> CNNs, Transformers, disadvantage for high-resolution chest X-ray modeling, due to lacking modeling power for very large context or quadratic time complexity-> S4, Mamba, U-Mamba, VisonMamba, linear time complexity and competitive modeling power as Transformers -> Motivation of bidirectional, Bi-directional Mamba, Multi-view, forward and backward blocks, two input patch concatenated chest X-rays. -> Dataset, Performance, Memory. Compared with performance in LDCT.
% Background\\
% Purpose\\
% Method\\
% Result\\
\keywords{Cardiovascular disease detection \and chest X-ray \and state space sequence models.}
\end{abstract}
\section{Introduction}
%Cardiovascular diseases (CVDs) stand as the foremost cause of mortality globally, posing a critical public health challenge~\cite{martin20242024}. Accurately predicting CVD risk through imaging techniques is essential in medical practice. This predictive ability is crucial for developing effective health management strategies, allowing for early intervention and potentially saving lives by slowing disease progression. 
Cardiovascular diseases (CVDs) are the leading cause of death worldwide, representing a significant public health challenge~\cite{martin20242024}. Accurate prediction of CVD risk using imaging techniques is vital in medical practice. Such predictive capability is key to devising effective health management strategies, enabling early intervention, and potentially saving lives by mitigating disease progression.
Previous studies have demonstrated that imaging features in computed tomography (CT) can help predict CVD risk~\cite{zeleznik2021deep,chao2021deep}. 
%Nonetheless, CT scanning is associated with significant radiation exposure, potentially leading to adverse health consequences for patients. This aspect limits its widespread acceptance and suitability for certain demographic groups. Conversely, chest X-ray is a more accessible imaging modality and emits substantially lower radiation doses than CT scans, making it a safer alternative for patients. This motivates us to investigate the viability of employing chest radiographs for CVD risk prediction.
Nonetheless, CT scanning entails notable radiation exposure, which may result in adverse health effects for patients. In contrast, chest X-ray stands as a more accessible imaging technique, emitting considerably lower levels of radiation compared to CT scans, thereby offering a safer option for patients. This rationale underpins our investigation into the feasibility of using chest X-ray for predicting CVD risk.

% readily available
% thereby presenting 
% \py{Start with a short paragraph about the background and then the problem you are dealing with.}

% \jj{(1) Background of cVD + xray replace CVD}

% \jj{(2) high resolution images in x-ray radiographs
% Merits: more detailed local textual information
% Problem: put challenge on the model to extract the global information.
% }

Recent strides in artificial intelligence, especially deep neural networks, have impressed the research community by their capability to capture effective imaging features for computer-aided diagnosis. Among all deep learning models~\cite{zhang2023neural,zhang2023revisiting,zhang2023spectral,zhang2023toward}, Convolutional Neural Networks (CNNs)~\cite{krizhevsky2012imagenet} and Transformers \cite{vaswani2017attention} have become two pivotal network architectures. CNNs (such as U-Net \cite{ronneberger2015u} and ResNet\cite{he2016deep}) employ shared-weight convolutional kernels to capture local features and is invariant to spatial translation. Transformers (such as ViTs \cite{dosovitskiy2020image} and Swin Transformer \cite{liu2021swin}) divide an image into a sequence of small patches and model multi-head self-attention between patches, encoding dense long-range dependencies. However, both CNNs and Transformers struggle to model very high resolution images due to the inherent locality of CNNs and the quadratic time complexity with respect to sequence length of Transformers. This problem is profound in chest X-ray disease detection since a typical chest radiographs contains more than a million pixels \cite{johnson2019mimic} and previous work \cite{haque2023effect} shows that down-scaling chest radiographs harms disease detection performance. The problem is even more challenging when using multiple views of chest X-ray imaging, which proves to bring performance benefits by previous studies \cite{rubin2018large,van2021multi,kim2023chexfusion}. Hence, an efficient architecture with long-range dependency modeling ability is demanded.

Recently, structured state space sequence models (S4) \cite{gu2021efficiently} have emerged as a promising class of network architectures for efficient sequence modeling. Mamba \cite{gu2023mamba} improves S4 with a selective mechanism, allowing the model to selectively forget or propagate information, achieving state-of-the-art on language and genomics modeling. With the diagonally structured state matrix and hardware-aware memory management, Mamba has linear time and memory complexity with respect to input sequence length. Inspired by these merits, recent works have transferred the Mamba network to the vision domain, such as U-Mamba \cite{ma2024u} for medical image segmentation and Vision Mamba (Vim) \cite{zhu2024vision} for natural image processing.

In this study, we propose a novel Bidirectional Image Mamba (BI-Mamba) model to efficiently process multi-view high-resolution chest radiographs for CVD risk prediction. To enrich the unidirectional Mamba with opposite directional information,
% \jj{what's the motivation of bi-direction processing}
% Bi-Mamba contains parallel forward and backward blocks in recurrent mode, encoding rich bidirectional long-range information into a classification token for CVD risk prediction. 
BI-Mamba utilizes parallel forward and backward blocks operating in recurrent mode. This structure captures extensive long-range information into a classification token for CVD risk prediction.
% Bi-Mamba stands out from prevalent Transformer-based architectures by offering linear scaling in image size, as opposed to Transformer's quadratic complexity.
BI-Mamba sets it apart from Transformers by its linear scaling with respect to image size, in contrast to Transformer's quadratic complexity. 
% -based architectures
% \jj{the sentences above are about the efficiency of Bi-Mamba inheritted from Mamba model, should we introduce the performance gain from the bi-directional design?} 
% Here, inspired by the recent success of the Mamba architecture \cite{gu2023mamba}, we present a novel Bidirectional Mamba (Bi-Mamba) to overcome the above limitations to efficiently process multi-view high-resolution chest X-ray radiographs.
% \cyan{to not only leverage the feature extraction power of foundation models but also effectively fuse the image features from both views of chest x-ray}. 
% \jj{Zefan, please add a paragraph here to briefly introduce the your method}
% \jj{move the details to the experiments section}
We use data from 10,395 subjects in the National Lung Screening Trail (NLST) \cite{national2011reduced} dataset to test BI-Mamba performance for CVD risk prediction in chest X-ray. Experimental results show that BI-Mamba achieves an Area under the Receiver Operating Characteristic Curve (AUROC) of 0.8243, outperforming ViT-S and ResNet-50 with comparable parameter size. BI-Mamba shows exciting memory efficiency with at least $30.5\%$ less allocated memory than ResNet-50 and ViT-S. Overall, our method shows promising performance compared with previous state of the art (0.8710) in CT \cite{chao2021deep}, unraveling the potential to use chest X-ray for CVD risk prediction. 
 % and $34.5\%$
% \jj{I suggest moving the implementation details in the Experiment section}

% % \py{What kind of experiments did you do? On what data? What conclusions can you draw?}

% \zf{X-ray high resolution, multi-view makes the resolution even higher $\rightarrow$ limitaions of ViTs and CNNs $\rightarrow$ advantage of state space models $\rightarrow$ our method: multi-view bidirectional}

% \blue{This paper has three major contributions. 
% \textbf{1)} To our best knowledge, this is the first work demonstrate the limitations of the existing models in the cvd diagonosis.
% \textbf{2)} Leveraging the efficient long-sequence processing capability of MamBa model, we proposed bi-direction multi-view Bi-Mambda (to be replaced by the model name) architecture for chest x-ray CVD diagnosis
% \textbf{3)} We conduct thorough empirical
% analyses to demonstrate the effectiveness of Bi-Mamba (to be replaced by the model name) in chest x-ray CVD diagnosis.}

% \section{Related Works}

% \py{Usually for MICCAI papers, we don't need a section for related works. But since this paper involves a few different aspects, I think it can be beneficial to have one.}

% \subsection{Fusion of features from multiple sources}

% Early fusion..

% Middle fusion...

% Late fusion...

% \subsection{State Space Modeling}

\section{Method}

\begin{figure}[t]
    \centering
    \includegraphics[width=\textwidth]{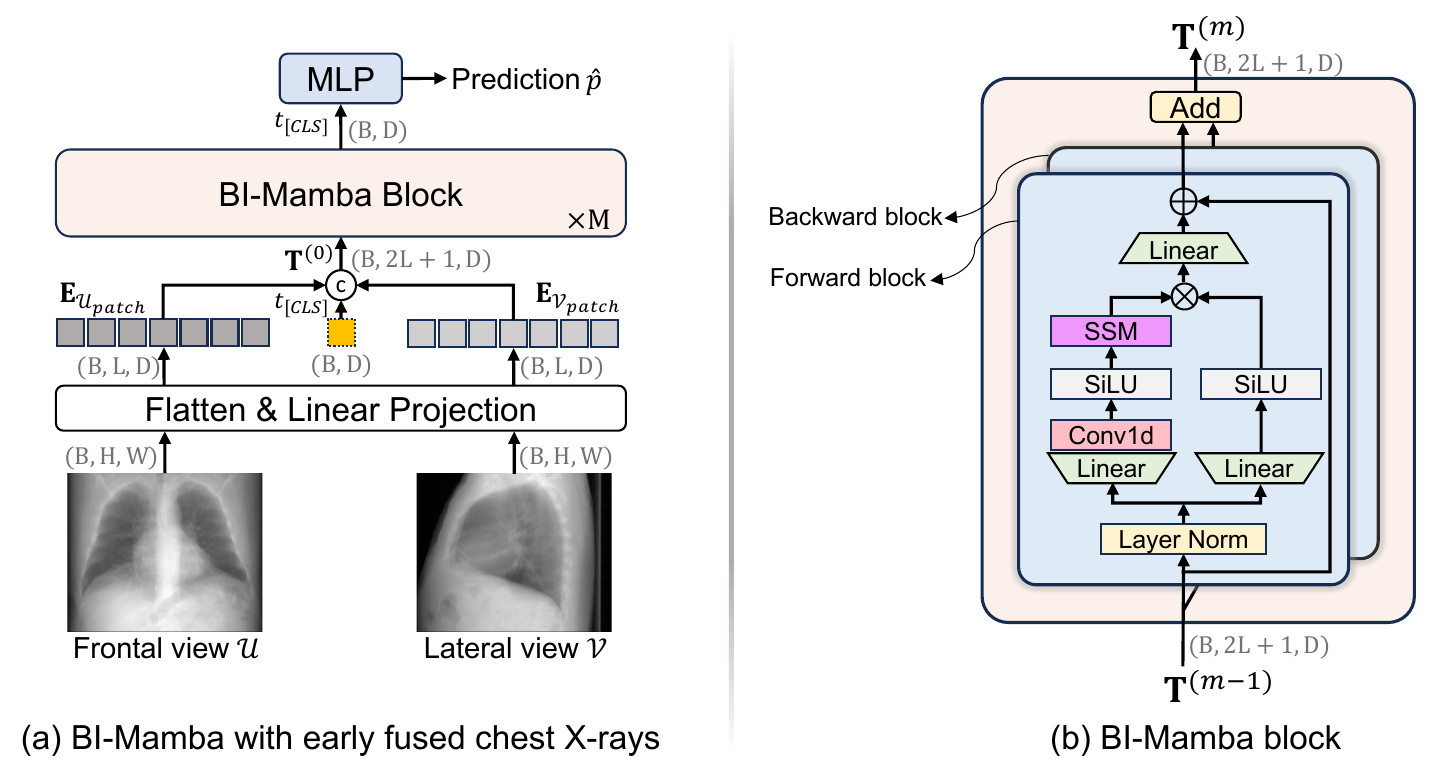}
    \caption{Network architecture of the BI-Mamba model.}
    \label{fig:framework}
\end{figure}

The proposed model for CVD risk prediction in chest X-ray consists of an input layer, a novel BI-Mamba model, and a multilayer perceptron (MLP) for output (Fig.\ref{fig:framework}(a)).
This section first describes the preliminaries of the state space models that are used for sequence modeling in BI-Mamba.
Then, we elaborate the details of the early fusion strategy and the specific operations inside the BI-Mamba block. 
Last, we analyze the computational complexity of the proposed model.
% for early fusion of the frontal view and lateral view images
% Our proposed Bi-Mamba method inherits the Mamba block from \cite{gu2023mamba} where state space models are used for sequence modeling, but designs parallel blocks that operates in forward and backward directions at each stage. We combine the frontal and lateral chest X-rays at the input layer and place a learnable classification token in the middle. 

% Thanks to the bidirectional characteristic and selective mechanism, the classification token can summarize rich long-range information from two views with severe domain shift during forward and backward passes. Since the SSM can be interpreted as a combination of CNNs and recurrent neural networks (RNNs), it enjoys linear scaling in sequence length, allowing for efficiently processing of high resolution images.

%\subsection{Preliminaries}
\subsection{Bidirectional Image Mamba (BI-Mamba)}

% \py{This section needs to be rewritten. Starting from an image input. Think about it, you are given an image. How would you start using Mamba for analysis. Go from there.}

\subsubsection{Preliminaries}
State space models represent dynamic physical systems by means of ordinary differential equations parameterized by $(\mathbf{A}, \mathbf{B}, \mathbf{C})$. 
% The continuous form of state space models is a function-to-function mapping $x(t) \in \mathbb{R} \mapsto y(t) \in \mathbb{R}$ through a hidden state $h(t) \in \mathbb{R}^N$. $\mathbf{A} \in \mathbb{R}^{N \times N}$ is the evolution parameter and $\mathbf{B} \in \mathbb{R}^{N\times1}$, $\mathbf{C} \in \mathbb{R}^{1\times N}$ are the projection parameters.
% \begin{equation}
% \begin{aligned}
% \label{eq:ssm}
% h'(t) &= \mathbf{A}h(t) + \mathbf{B}x(t), \\
% y(t) &= \mathbf{C}h(t).
% \end{aligned}
% \end{equation}
Mamba \cite{gu2023mamba} uses a discrete form of the continuous state space model. The discretization is achieved by transforming differential integral to summation across a time step $\mathbf{\Delta}$. With this discretization, the state space model can be written in a recurrent form, where the output $y \in \mathbb{R}$ at $t$ is determined by the hidden state $h \in \mathbb{R}^{N}$ at $t-1$ and the input $x \in \mathbb{R}$ at $t$:
\begin{equation}
\label{eq:disc_ssm}
\begin{aligned}
h(t) &= \overline{\mathbf{A}}h(t-1) + \overline{\mathbf{B}}x(t), \\
y(t) &= \mathbf{C}h(t),
\end{aligned}
\end{equation}
where $\overline{\mathbf{A}} \in \mathbb{R}^{N\times N} = f_\mathbf{A}(\mathbf{\Delta}, \mathbf{A})$, $\overline{\mathbf{B}} \in \mathbb{R}^{N \times 1} = f_\mathbf{B}(\mathbf{\Delta}, \mathbf{A}, \mathbf{B})$, and $\mathbb{\mathbf{C}} \in \mathbb{R}^{1 \times N}$ are state matrices. The pair $(f_\mathbf{A}, f_\mathbf{B})$ is a discretization rule. The above equation represents the core operation inside the SSM module in Fig.\ref{fig:framework}(b).

% \py{Why A and B have bars on top? B and C seem to be vectors? if that is the case, they should be in lower case bold letters. More importantly, what do they represent? How is that related to our problem? Any intuitive illustration?} \zf{Capital B and C are conventionally used in the field of state space models.}
% $\overline{\mathbf{A}} \in \mathbb{R}^{N\times N}$, $\overline{\mathbf{B}} \in \mathbb{R}^{N \times 1}$, and $\mathbb{\mathbf{C}} \in \mathbb{R}^{1 \times N}$ are state matrices

\subsubsection{BI-Mamba Block}
BI-Mamba belongs to the family of state space sequence models (SSMs) that receive token sequences as input. Given a gray-scale medical image $\mathcal{U} \in \mathbb{R}^{H\times W}$ as input, it has to first rearrange into a sequence of small patches $\mathcal{U}_{patch} \in \mathbb{R}^{J\times (P\times P)}$, where $(H, W)$ is the image size, $J$ is the number of patches, and $(P, P)$ is the patch size. 
A linear projection layer then maps $\mathcal{U}_{patch}$ to patch embeddings $\mathbf{E}_{\mathcal{U}_{patch}} \in \mathbb{R}^{J\times D}$, where $D$ is the embedding dimension.
A learnable classification token $\mathbf{t}_{[CLS]}$ is then initialized and placed at the middle of $\mathbf{E}_{\mathcal{U}_{patch}}$. Different from the position-agnostic $\mathbf{t}_{[CLS]}$ in ViTs \cite{dosovitskiy2020image}, a middle $\mathbf{t}_{[CLS]}$ proves to be better than left-ended one in Mamba \cite{zhu2024vision}. The input to BI-Mamba is denoted by:
\begin{equation}
\mathbf{T}^{(0)} = [\mathbf{E}_{\mathcal{U}_{patch}}^{(1)}, ..., \mathbf{t}_{[CLS]}, ..., \mathbf{E}_{\mathcal{U}_{patch}}^{(J)}] + \mathbf{E}_{pos},   
\end{equation}
where $\mathbf{T}^{(0)}, \mathbf{E}_{pos} \in \mathbb{R}^{L\times D}$ are the input sequence and positional embeddings and $L=J+1$ is the sequence length.

A BI-Mamba block (Fig.\ref{fig:framework}(b)) consists of parallel forward and backward blocks that encode the input sequence in opposite directions, producing representations conditioning on both left and right context. 
% \py{Given an image, what does this mean? Instead of encoding starting from the top left, it also goes from bottom right? How would this help image analysis?}
Inside the forward or backward block, $\mathbf{T}^{(m-1)}$ is first normalized and projected to two variables $x, z \in \mathbb{R}^{L\times E}$ in expanded state dimension $E$. $E$ is multiple times larger than $D$. $x$ is the input to the SSM module, representing the input sequence. $z$ is the variable for gating dependent on the input sequence. 
% \py{What do x and z represent in terms of our task?} $E$ is the expanded state dimension, usually double the dimension of the embedding dimension $D$. \py{Scalar values like the dimension should be in lower cases.} 
\begin{equation}
\begin{aligned}
&\mathbf{T}'^{(m-1)} = \textbf{Norm}(\mathbf{T}^{(m-1)}), \\
&x = \textbf{Linear}^x(\mathbf{T}'^{(m-1)}), \\
&z = \textbf{Linear}^z(\mathbf{T}'^{(m-1)}).
\end{aligned}
\end{equation}
% \py{What is x? Image patches? Spell it out.} 
The projected input sequence $x$ then passes through a 1D convolution and activation unit $x'_d = \textbf{SiLU}(\textbf{Conv1d}_d(x))$ to create non-linearity, where $x'_d\in\mathbb{R}^{L \times E}$. The subscript $d\in\{forward, backward\}$ is used to indicate features or operation units specific to a directional block. 

A key property of Mamba is that the SSM parameters are input-dependent, as apposed to time- (\textit{i.e}., patch position) and input-invariant, making it powerful to model images with large appearance shift. BI-Mamba makes it even more powerful by letting the SSM parameters $(\mathbf{\Delta}_d, \mathbf{A}_d, \mathbf{B}_d, \mathbf{C}_d)$ condition on context direction. The SSM parameters are generated by the linear projection of $x'_d$ and initialized parameters.
% \begin{equation}
% \begin{aligned}
% \mathbf{A}_d &= \textbf{Param}^\mathbf{A}_d, \\
% \mathbf{B}_d &= \textbf{Linear}^B_d(x'_d), \\
% \mathbf{C}_d &= \textbf{Linear}^C_d(x'_d), \\
% \mathbf{\Delta}_d = \textbf{SoftPlus}&(\textbf{Linear}^\mathbf{\Delta}_d(x'_d)+\textbf{Param}^\mathbf{\Delta}_d), \\
% \end{aligned}
% \end{equation}
% where $\mathbf{A}_d \in \mathbb{R}^{E\times N}$, $\mathbf{B}_d \in \mathbb{R}^{B\times L' \times N}$, $\mathbf{C}_d \in \mathbb{R}^{B \times L' \times N}$, and $\mathbf{\Delta} \in \mathbb{R}^{B\times L' \times E}$. The $\textbf{SoftPlus}$ function ensures positive time step values. 
Discretizing $(\mathbf{A}_d, \mathbf{B}_d)$ given $\mathbf{\Delta}_d$ to obtain the form in Eq.~(\ref{eq:disc_ssm}) is achieved by a simple multiplication discretization rule $(f_\mathbf{A}, f_\mathbf{B})$.
\begin{equation}
\overline{\mathbf{A}}_d = \mathbf{\Delta}_d \otimes \mathbf{A}_d, \overline{\mathbf{B}}_d = \mathbf{\Delta}_d \otimes \mathbf{B}_d,
\end{equation}
where $\mathbf{\Delta}_d \in \mathbb{R}^{L \times E}$, $\overline{\mathbf{A}}_d, \overline{\mathbf{B}}_d \in \mathbb{R}^{L \times E \times N}$, and $\mathbf{C}_d \in \mathbb{R}^{L \times N}$. $N$ is the latent dimension in SSM, typically set to a small value (\textit{e.g}. 16). The SSM parameterized by $(\overline{\mathbf{A}}_d, \overline{\mathbf{B}}_d, \mathbf{C}_d)$ recurrently processes the input variable $x'_d$ along the forward or backward direction, producing the output sequence $y_d \in \mathbb{R}^{L \times E}$, which is then filtered by $z$ via element-wise product. Then, the $\textbf{Linear}^\mathbf{T}$ maps $y_d$ back to the embedding dimension $D$.
\begin{equation}
\begin{aligned}
&y_d = \textbf{SSM}_d(x'_d) \otimes \textbf{SiLU}(z), \\
&\mathbf{T}^{(m)}_d = \textbf{Linear}^{\mathbf{T}}(y_d ) + \mathbf{T}^{(m-1)}.
\end{aligned}
\end{equation}
The final output $\mathbf{T}^{(m)}$ is the summation of $\mathbf{T}^{(m)}_{forward}$ and $\mathbf{T}^{(m)}_{backward}$, and hence jointly encapsulates both left-to-right and right-to-left dependencies. The classification token from the output of the last BI-Mamba block $\mathbf{T}^{(M)}$ is projected by a MLP and sigmoid function to produce the prediction $\hat{p}$, which is then supervised by a cross-enropy loss.

% \py{My overall feeling of the above description is that it is very generic. Your readers are MIC people. They are intereseted in how to fit images into this process. Think about the ViT paper, does it just talk about attention and transformer blocks? It is about how you apply Mamba to image analysis. Tell people how Mamba works for image data. Explain why Bi-Mamba is even better.}

% \begin{equation}
% \begin{aligned}
% \overline{\mathbf{A}} &= \exp(\mathbf{\Delta}\mathbf{A}), \\
% \overline{\mathbf{B}} &= (\mathbf{\Delta}\mathbf{A})(\exp(\mathbf{\Delta}\mathbf{A}) - \mathbf{I}) \cdot \mathbf{\Delta}\mathbf{B}.
% \end{aligned}
% \end{equation}

\subsection{BI-Mamba for Multi-View Image Analysis}

% \py{Part of this paragraph should actually be moved earlier. Talk about single image input first. Then it will be easy to extend to multiple images here.}
%\subsubsection{Input patch concatenated Chest X-rays}
% Bi-Mamba is a sequence model that receives a sequence of patch tokens as input. The frontal and lateral chest X-rays $u, v \in \mathbb{R}^{B\times H\times W}$ are first rearranged into a sequence of two dimensional patches $u_{patch}, v_{patch} \in \mathbb{R}^{B\times L}$, where $B$ is the batch size and $L = \frac{H}{P} \times \frac{W}{P}$ is the number of patches, and $P$ is the patch size. 
% \py{Don't use B for two reasons. 1. It is unnecessary to mention batch size. 2. B has been used before when you introduce SSM.} 
% A linear projection layer then maps $u_{patch}, v_{patch}$ into patch embeddings $\mathbf{E}_{u_{patch}}, \mathbf{E}_{v_{patch}} \in \mathbb{R}^{B\times L\times D }$, where $D$ is the embedding dimension. 
% \py{Justify the fusion or at least mention other possibilities of fusion, e.g. early, mid, late, etc.}
Previous studies have applied different fusion strategies to integrate multi-view chest X-rays for disease detection, such as middle fusion \cite{van2021multi} and late fusion \cite{rubin2018large,kim2023chexfusion}. Inspired by the efficient linear complexity in sequence length of BI-Mamba, we propose a novel early input patch concatenation strategy for multi-view combination. This early fusion manner allows BI-Mamba to sufficiently model the synergy between chest X-ray views. The input sequence $\mathbf{T}^{(0)}$ is rewritten as:
% As in Vision Transformers, a learnable classification token $\mathbf{t}_{[CLS]} \in \mathbb{R}^{B \times D}$ is initialized to aggregate global information for classification. $\mathbf{E}_{u_{patch}}, \mathbf{E}_{v_{patch}}$ are concatenated along the length dimension and $\mathbf{t}_{[CLS]}$ is placed in their middle. Finally, a positional embedding $\mathbf{E}_{pos} \in \mathbb{R}^{B \times L'}$ is added to the early fused sequence $\mathbf{T}^{(0)} \in \mathbb{R}^{B\times L'\times D}$, where $L'=2L+1$.
\begin{equation}
\mathbf{T}^{(0)} = [\mathbf{E}_{\mathcal{U}_{patch}}^{(1)}, ..., \mathbf{E}_{\mathcal{U}_{patch}}^{(L)}, \mathbf{t}_{[CLS]}, \mathbf{E}_{\mathcal{V}_{patch}}^{(1)}, ..., \mathbf{E}_{\mathcal{V}_{patch}}^{(L)}] + \mathbf{E}_{pos},
\end{equation}
where $\mathbf{E}_{\mathcal{U}_{patch}}, \mathbf{E}_{\mathcal{V}_{patch}}$ denote the patch embeddings of the frontal view $\mathcal{U}$ and lateral view $\mathcal{V}$ respectively and $\mathbf{T}^{(0)} \in \mathbb{R}^{(2L+1) \times D}$.
% $\mathbf{T}^{(m-1)}$ denotes the input to the $m$-th Bi-Mamba block to produce an output $\mathbf{T}^{(m)}$.

\subsection{Computational complexity}
\label{sec:bigo}
% mamba block
% The embedding dimension of the proposed model is $D$, the expanded state dimension is $E$, and the SSM latent dimension is $N$.
The state space model independently maps each channel of the input $x \in \mathbb{R}^{L \times E}$ to the output $y \in \mathbb{R}^{L \times E}$ through a higher dimensional state $h \in \mathbb{R}^{N}$. This results in $EN$ number of effective states per input. With $\mathbf{A} \in \mathbb{R}^{N\times N}$ representing by diagonal structure \cite{gu2021efficiently}, computing them over an input of length $L$ requires $O(LEN)$ time. In contrast, self-attention in ViTs has a time complexity of $O(L^2D)$. Since self-attention has quadratic scaling to sequence length, it is much more inefficient when handling long sequences. Following \cite{zhu2024vision}, we set the number of blocks $M$ to 24, embedding dimension $D$ to 384, expanded state dimension $E$ to 768, and SSM latent dimension to $16$.
% \py{The current analysis doesn't provide much context to understand the complexity. For example, you should provide the complexity of ViT and CNN, given similar settings.}
 % per directional block
 % and batch size $B$ 
% \noindent Displayed equations are centered and set on a separate
% line.
% \begin{equation}
% x + y = z
% \end{equation}

\section{Experiments and Results}
\subsection{Datasets}

This study uses the National Lung Screening Trail (NLST) dataset \cite{national2011reduced} in experiments. We obtained the same low dose CT scans from NLST and the same CVD risk labels as the state-of-the-art method for CVD prediction in CT \cite{chao2021deep}. We then simulated chest radiographs via parallel projection through CT volumes \cite{moturu2018creation} to create chest X-ray images for direct performance comparison with the model on CT images. We simulated chest radiographs in both frontal and lateral views.\footnote{Details of chest X-ray simulation are shown in Supplementary Material.} 
% by accumulating voxel attenuation coefficients along anterior-posterior paths and left-right paths, respectively.

In total, the dataset has 33,413 low dose CT scans from 10,395 subjects. From each CT scan, we generated frontal and lateral chest radiographs, resulting in 66,826 images. Subjects are categorized into high and low CVD risk groups according to their exam reports and causes of death for deceased subjects. Specifically, a subject is deemed having high CVD risk when there is cardiovascular abnormality reported in the exam report or the subject died of CVD according to the ICD-9 code. A subject is considered normal when the subject has no CVD-related medical history and no reported cardiovascular abnormality in CT, and did not die of circulatory system diseases. The numbers of subjects with high and low CVD risk are 2,962 and 7,433 respectively. The chest X-ray data is divided into training (70\%, 7,268 subjects), validation (10\%, 1,042 subjects), and test splits (20\%, 2,085 subjects) for model development and evaluation. 

\subsection{Experimental details}
\begin{figure}[t]
\centering
\begin{subfigure}{.41\textwidth}
  \centering
  \includegraphics[width=\linewidth]{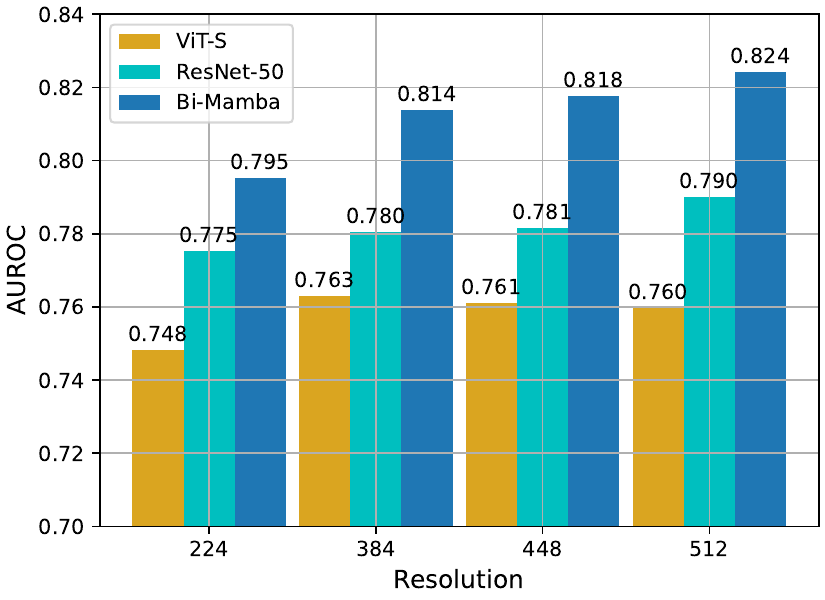}
  \caption{}
  \label{fig:sub1}
\end{subfigure}%
\begin{subfigure}{.5\textwidth}
  \centering
  \includegraphics[width=.75\linewidth]{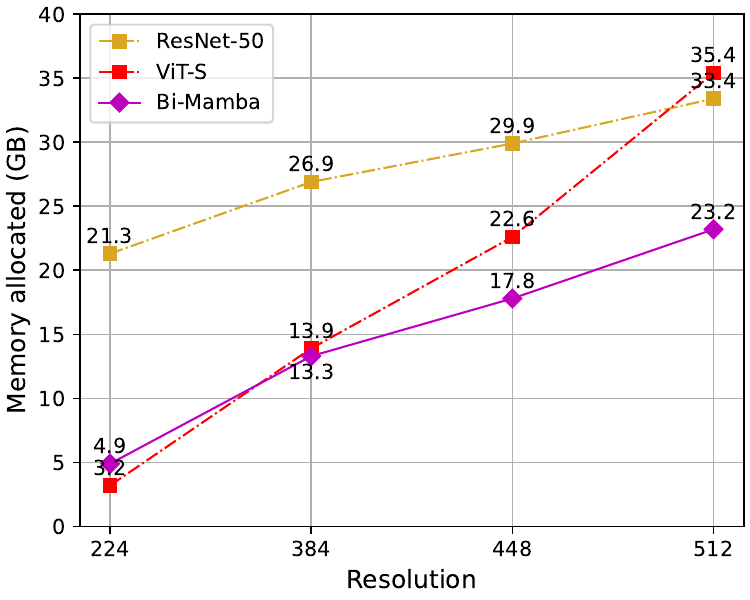}
  \caption{}
  \label{fig:sub2}
\end{subfigure}
\caption{Results on different image resolution and comparison of memory footprint at training time.}
\label{fig:res}
\end{figure}

We set the model input chest X-ray size to $512$ by default. For results in Fig.\ref{fig:res}(a), we vary image size to $448$, $384$, and $224$. 
% \py{Fig. 2a shows you have several different image sizes. Here it reads like you have only one fixed size. Readers will be confused.} 
For sequence models (ViT-S and BI-Mamba), we use a patch size 16. During model training, we apply random crop (aspect ratio ranging from (0.75, 1.3)) and horizontal flip as image augmentation. We initialized our BI-Mamba model with the weights released by the Vision Mamba team \cite{zhu2024vision}. ViTs were initialized with weights from DINOv2 \cite{oquab2023dinov2}. ResNet-50 was initialized with weights pretained on ImageNet. We set the initial learning rate to 5e-6, weight decay to 1e-8, and batch size to 24. We used the AdamW optimizer and cosine learning rate scheduler. The number of training epochs is 30. Models are trained on a NVIDIA DGX-1 server with 8x A100 GPUs. At the inference time, we compute the area under the receiver operating characteristic (AUROC) curve in the test split to evaluate classification performance. Source code is publicly available at: \url{https://github.com/RPIDIAL/BI-Mamba}.

% \py{size ratio??}
% Benchmarks: single vs multimodal approaches in neuro
% non-attention fusion mechanisms
% horizontal flip, learning rate 5e-6, weight decay 13-8, batch size 24, epoch 30, patch szie 16x16, 8 A100 GPUs, 
% Insert figures for clustering results
% \py{AUROC values or ROC curves?}

\subsection{Results of CVD Risk Prediction}

We compared our BI-Mamba model with other methods using different network architectures and view combination strategies (Table~\ref{tab:fusion}). The benchmarking networks include ResNet-50 and ViT-S.  ResNet-50 combines features from two chest X-ray views after the global average pooling in each branch~\cite{rubin2018large}.
 % and projects them to classification logits
The ResNet-50 with cross attention \cite{van2021multi} includes a cross attention block after the third block of ResNet-50.
 % to combine view-specific representations
ViT-S and BI-Mamba both use a $\texttt{[CLS]}$ token as their final output. They combine the $\texttt{[CLS]}$ tokens from the last layer outputs of the two views for prediction. ViT-S and BI-Mamba with input patch concatenation directly combine two patch sequences at the input layer and have only one \texttt{[CLS]} token. ViT-S conventionally places the \texttt{[CLS]} token at the left-end. BI-Mamba puts it at the middle between the two input patch sequences. 
Table \ref{tab:fusion} shows that BI-Mamba with input patch concatenation attains a 0.8243 AUROC and outperforms all the other networks and combination strategies.
Under the same combination strategy, ViT-S attains a 0.7597 AUROC, significantly inferior to BI-Mamba's performance ($p<0.05$). This result can be attributed to their different mechanisms for sequence modeling. 
% Bi-Mamba uses bidirectional recurrent formulation and a selective mechanism to forget or propagate information. 
A middle $\texttt{[CLS]}$ in BI-Mamba independently aggregate information from one view in recurrent propagation. Conversely, ViT-S uses self-attention, which may be deficient to model sequences from two views, since there are not always strong correspondences to attend to between the query from one view and keys and values from the other view. 

Besides, we compared BI-Mamba with the state of the art method Tri-2D network \cite{chao2021deep} for CVD risk prediction. Tri-2D attains a 0.8710 AUROC in low-dose CT. BI-Mamba presents a promising result of 0.8243 in chest X-rays.
% Bi-Mamba with input patch concatenation also outperforms ResNet-50 and ViT-S with the output/$\texttt{[CLS]}$ concatenation and cross attention, probably owing to its sufficient encoding of two-view information thanks to the combination at the earliest stage possible.

% , without interference of the other view.
 % reduces the maximum path length to $O(\frac{L}{2})$ to any patch tokens
% ResNet+CA stands for ResNet-50 plus cross-view attention, where the cross-view attention resides after the third block of the ResNet-50.

% \py{Add a new column of image encoder. The first two columns should be clarified. Right now network architecture and specific fusion methods are mixed together. Use multirow to help format.}

% \py{Add stat test information here.}

% \toprule
% Fusion      & Architecture   & \# params & AUROC  & $p$-value\\ \midrule
% Late & ResNet-50 & 25M          & 0.7900 & 2e-5 \\
%             & ViT-S     & 21M          & 0.7877 & 4e-5   \\
%             & Bi-Mamba  & 25M          & 0.8106 & 0.1297 \\ \midrule
% Middle & ResNet+CA & 29M        & 0.7914 & 4e-5 \\ \midrule
% Early  & ViT-S     & 21M        & 0.7597 & 5e-8 \\
%               & Bi-Mamba  & 25M        & 0.8243 & \\
% \bottomrule      

\begin{table}[t]
\setlength{\tabcolsep}{8pt}
\centering
\caption{Results of different network architectures and view combination strategies. $p$-value is computed with DeLong's algorithm \cite{sun2014fast} that compares the statistical difference between AUROC values.}
\label{tab:fusion}
\begin{tabular}{c c c c c}
\toprule
Architecture & View Combination   & \# params & AUROC  & $p$-value\\ \hline
\multirow{2}{*}{ResNet-50}    & Output concat. \cite{rubin2018large} & 25M       & 0.7900 & 2e-5 \\ 
             & Cross Attention \cite{van2021multi} & 29M       & 0.7914 & 4e-5 \\ \hline
\multirow{2}{*}{ViT-S}        & \texttt{[CLS]} token concat.  & 21M       & 0.7877 & 4e-5   \\
        & Input patch concat. & 21M       & 0.7597 & 5e-8 \\ \hline
\multirow{2}{*}{BI-Mamba}     & \texttt{[CLS]} token concat.  & 25M       & 0.8106 & 0.1297 \\
     & Input patch concat. & 25M       & \textbf{0.8243} & - \\
\toprule      
\end{tabular}
\end{table}

% \begin{figure}
% \centering
% \begin{minipage}{0.5\textwidth}
%     \centering
%     \includegraphics[width=0.9\linewidth]{auroc_size.pdf}
%     \captionof{figure}{Results on early fusion strategy on Bi-Mamba with different image sizes}
%     \label{fig:auroc}
% \end{minipage}% \hfill
% \begin{minipage}{0.5\textwidth}
%     \centering
%     \includegraphics[width=0.9\linewidth]{memory.pdf}
%     \captionof{figure}{Benchmarking computational efficiency on ResNet-50, ViT-S, and Bi-Mamba.}
%     \label{fig:memory}
% \end{minipage}
% \end{figure}

% \begin{figure}
%     \centering
%     \includegraphics[width=0.7\textwidth]{auroc_size.pdf}
%     \caption{Results on early fusion strategy on Vim with different image sizes.}
%     \label{fig:enter-label}
% \end{figure}

% \begin{table}
% \centering
% \caption{Benchmarking computational efficiency on early fusion strategy on Vim and ViT-S.}
% \label{tab:eff}
% \begin{tabular}{c | c | c | c | c | c | c}
% \hline
% Network & \# params & Image size  & Patch size  & Stride & Speed & Memory  \\ \hline
%  Vim    & 25M       & 512         & 16          & 8      &       &         \\
%  Vim    & 25M       & 512         & 16          & 16      &       &         \\
%  ViT-S  & 22M       & 512         & 14          & 14      &       &         \\
% \hline
% \end{tabular}
% \end{table}

\begin{table}[t]
\setlength{\tabcolsep}{3.5pt}
\centering
\caption{Risk prediction results using each view individually.}
% \py{Please specify `Method' to be something more meaningful.} \zf{I add the initialization weight column.}
\label{tab:baseline}
\begin{tabular}{c c c c c c }
\hline
\multirow{2}{*}[-2pt]{Architecture}  & \multirow{2}{*}[-2pt]{\# params} & \multirow{2}{*}[-2pt]{embed dim} & \multirow{2}{*}[-2pt]{init. weights} & \multicolumn{2}{c}{AUROC}  \\ \cmidrule(lr){5-6} 
 & & & & Frontal & Lateral \\ \midrule
ResNet-50     & 25M          & -         & sup. pretrain.     & 0.7831  & 0.7800   \\ \hline
ViT-S         & 21M          & 384       & DINOv2 pretrain.  & 0.7712  & 0.7680   \\
ViT-B         & 86M          & 768       & DINOv2 pretrain.  & 0.7649  & 0.7660   \\
ViT-L         & 300M         & 1024      & DINOv2 pretrain.  & 0.7770  & 0.7816   \\ \hline
BI-Mamba      & 25M          & 384       & sup. pretrain.     & \textbf{0.7924}  & \textbf{0.7946}   \\
\hline 
\end{tabular}
\end{table}

\subsection{Ablation studies}
\subsubsection{Results on different image resolution}
To investigate whether image resolution is a bottleneck of CVD risk prediction in chest X-ray, we gradually increase image size from 224 to 512. We apply output concatenation to ResNet-50 and input patch concatenation to ViT-S and BI-Mamba in experiments. As shown in Fig. \ref{fig:res}(a), increasing image size steadily improves BI-Mamba's performance, leading to an AUROC margin of 2.9\% between image size 512 and 224. We expect the tendency to extrapolate with higher image resolution, which we will validate in our future work. 
% ResNet-50 and ViT-S show lower performance than Bi-Mamba across image resolution for CVD risk prediction in chest X-ray.

\subsubsection{Analysis of computational efficiency}
Computational efficiency is an inevitable factor to consider when input size scales. BI-Mamba and ViTs have computational complexities of $O(L^2D)$ and $O(LEN)$ respectively. ViTs have quadratic time complexity with respect to input length, while BI-Mamba has linear time complexity, making it more efficient with high-definition inputs. This is evidenced by 34.5\% less allocated memory for BI-Mamba at image resolution 512 (Fig.\ref{fig:res}(b)).
% In Bi-Mamba, the state space model has a time complexity of $O(LEN)$. 
% The self-attention mechanism in 
% computes the dot product of a query patch with all patches in a sequence, resulting in $L\times D$ computational cost. Computing this self-attention over the sequence length of $L$ and batch size of $B$ makes
 % as analyzed in Section \ref{sec:bigo}

\subsubsection{Risk prediction results using individual views}
This section provides single-view results to benchmark the benefits of multi-view chest radiographs for CVD risk prediction. We load ImageNet supervised pretraining weights for ResNet-50 and BI-Mamba, and perform fine-tuning. We load DINOv2 general purpose features for ViT-S, ViT-B, and ViT-L, and perform linear probing. 
% \py{How about ViT-S in Table 1? If that is also DINO pretrained, it should have been mentioned there.}
Results are shown in Table \ref{tab:baseline}
% Please try to avoid rasterized images for line-art diagrams and
% schemas. Whenever possible, use vector graphics instead (see
% Fig.~\ref{fig1}).

% \begin{figure}
% \includegraphics[width=\textwidth]{fig1.eps}
% \caption{A figure caption is always placed below the illustration.
% Please note that short captions are centered, while long ones are
% justified by the macro package automatically.} \label{fig1}
% \end{figure}

\section{Conclusions}
We propose a novel Bidirectional Image Mamba (BI-Mamba) model for cardiovascular disease (CVD) risk prediction in multi-view chest X-rays. BI-Mamba contains parallel forward and backward blocks in recurrent mode to encode rich long-range dependencies of two input concatenated chest X-rays. BI-Mamba shows superior modeling power than ViT-S and ResNet-50, while it only has linear time complexity and saves at least 30.5\% training-time GPU memory. Compared with previous state of the art in CT (0.8710), BI-Mamba attains a promising 0.8243 AUROC, unlocking the gate towards a low-dose, inexpensive, and accurate computed-aided CVD risk prediction technique in chest X-ray. 
% In future works, thanks to Bi-Mamba efficient and effective properties, extend it to model real-world very high resolution chest X-ray imaging \cite{johnson2019mimic,national2011reduced} is plausible and promising.
% when trained on high resolution chest X-rays
% in multi-view chest X-rays for CVD risk prediction
% Photon counting x-ray, very high resolution

\bibliographystyle{splncs04}
\bibliography{refs}

\end{document}